\documentclass{ws-procs975x65}

\usepackage{epsfig}

\newcommand{\beqs}{\begin{equation*}}
\newcommand{\beq}{\begin{equation}}

\newcommand{\eeqs}{\end{equation*}}
\newcommand{\eeq}{\end{equation}}

\newcommand{\beqas}{\begin{eqnarray*}}
\newcommand{\beqa}{\begin{eqnarray}}

\newcommand{\eeqas}{\end{eqnarray*}}
\newcommand{\eeqa}{\end{eqnarray}}

\newcommand{\eq}[2]{\begin{equation} #1 \label{#2} \end{equation}}

\newcommand{\eps}{\varepsilon}

\newcommand{\om}{\omega}

\newcommand{\blist}{\begin{itemize}}

\newcommand{\elist}{\end{itemize}}

\providecommand{\href}[2]{#2}

\DeclareFontFamily{OT1}{rsfs}{}
\DeclareFontShape{OT1}{rsfs}{m}{n}{ <-7> rsfs5 <7-10> rsfs7 <10->rsfs10}{} 
\DeclareMathAlphabet{\mycal}{OT1}{rsfs}{m}{n}
\newcommand{\scri}{{\mycal I}}

\begin{document}

\title{Deformations of the Schwarzschild Black Hole}

\author{D. Grumiller}

\address{Institut f.\ Theoretische Physik, TU Wien, Wiedner Hauptstr.\ 8-10/136,\\A-1040 Vienna, Austria, Europe and\\Institut f.\ Theoretische Physik, Universit\"at Leipzig, Augustusplatz 10,\\D-04109 Leipzig, Germany, Europe\\E-mail: grumil@hep.itp.tuwien.ac.at}   

\maketitle

\abstracts{ 
Due to its large number of symmetries the Schwarzschild Black Hole can be described by a specific two-dimensional dilaton gravity model. After reviewing classical, semi-classical and quantum properties and a brief discussion of virtual black holes deformations are studied: the first part is devoted to deformations of the Lorentz-symmetry, the second part to dynamical deformations and its role for the long time evaporation of the Schwarzschild Black Hole.
}



\enlargethispage{2cm}

\section{Introduction}

Although the focus will be on the Schwarzschild Black Hole (SS BH) it is emphasized that most considerations below are valid for generic dilaton gravity models in two dimensions (2d). In this context the review\cite{Grumiller:2002nm} may be consulted. The main idea behind a study of the SS BH as a 2d effective theory is to gain insight into the otherwise more difficult accessible realm of quantum gravity by exploiting advantageous features of dilaton gravity in its first order formulation.

\section{Classical SS BH}

The SS BH may be studied at the level of 2d dilaton gravity by means of spherical reduction\cite{Berger:1972pg,Kuchar:1994zk}. 
Especially the first order\footnote{The following notation\cite{Grumiller:2002nm} is used: $e^a$ is the zweibein one-form, $\epsilon = e^+\wedge e^-$ is the volume two-form. The one-form $\omega$ represents the  spin-connection $\om^a{}_b=\eps^a{}_b\om$ with  the totally antisymmetric Levi-Civit{\'a} symbol $\eps_{ab}$ ($\eps_{01}=+1$). With the flat metric $\eta_{ab}$ in light-cone coordinates ($\eta_{+-}=1=\eta_{-+}$, $\eta_{++}=0=\eta_{--}$) the first (``torsion'') term of (\ref{cs:1}) is given by $X_a(D\wedge e)^a =
\eta_{ab}X^b(D\wedge e)^a =X^+(d-\omega)\wedge e^- + X^-(d+\omega)\wedge e^+$. Signs and factors of the Hodge-$\ast$ operation are defined by $\ast\epsilon=1$. The auxiliary fields $X,X^a$ can be interpreted as Lagrange multipliers for geometric curvature and torsion, respectively. $X$ usually is referred to as ``dilaton'' and can be interpreted also as surface area for the SS BH; $X^\pm$ correspond to the expansion spin coefficients $\rho,\rho'$ (both are real).} formulation\cite{Schaller:1994es} 
\eq{
L={\rm const} \cdot \int_{\mathcal{M}_2} \left[X_a (D\wedge e)^a +Xd\wedge\omega + \epsilon {\mathcal V} (X,X^aX_a) \right]\,,
}{cs:1}
turned out to be very useful already classically\cite{Klosch:1996fi} 
because it allowed the construction of the corresponding Carter-Penrose (CP) diagram with particular ease. For the SS BH the potential reads ${\mathcal V} = -X^+X^-/(2X)-\rm const.$

\section{Semi-classical SS BH}

Particle creation due to the Hawking effect exists also in 2d models. For technical details cf.\ refs.\cite{Kummer:1999zy,Grumiller:2002nm} and references therein.

\section{Quantum SS BH}

Kucha{\v{r}} has shown\cite{Kuchar:1994zk} that the mass of the SS BH obeys a Schr\"odinger equation. At the level of dilaton gravity this can be argued as follows: since---in the absence of matter and test particles---the only physical observable is the so-called Casimir function (which for the SS BH is essentially given by the ADM mass) any Schr\"odinger functional can only depend on it (and its conjugate variable, the extrinsic time).

\section{Virtual SS BH}

\begin{figure}
\centering
\epsfig{file=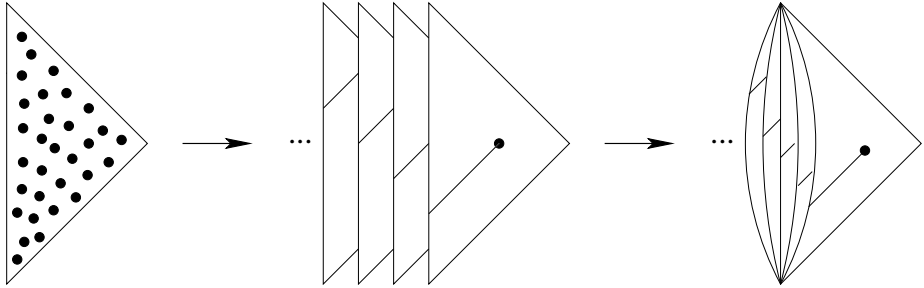,width=0.96\linewidth}
\caption{Constructing the CP diagram of all VBHs}
\label{fig:2}
\end{figure}

Instead of fixing a geometric background and quantizing matter on it, it is possible to integrate out geometry first\cite{Kummer:1998zs} and to perform a perturbative expansion in the matter degrees of freedom afterwards\cite{Kummer:1998zs,Grumiller:2000ah}. 
If this is done to lowest order for scattering of s-waves of massless scalars\cite{Fischer:2001vz} virtual BHs appear at an intermediate stage\cite{Fischer:2001vz,Grumiller:2001rg}. 
Due to our choice of ``temporal gauge'' (which was a technical pre-requisite for a successful path integration of geometry) the ensuing line-element has (outgoing) Sachs-Bondi form
\begin{equation}
(ds)^2 = 2 dr du + \left(1 - \frac{2m(u,r)}{r} - a(u,r) r + d(u,r)\right) (du)^2\,,
\label{ds2}
\end{equation}
with $m$, $a$ and $d$ being localized\footnote{The localization of ``mass'' 
and ``Rindler acceleration'' on a light-like cut is not an artifact of an accidental gauge choice, but has a physical interpretation in terms of the Ricci-scalar. Certain parallels to Hawking's Euclidean VBHs can be observed, but also essential differences. The main one is our Minkowski signature which we deem to be a positive feature.} 
on the cut $u=u_0$ with compact support $r<r_0$. These quantities depend on the second set of coordinates $u_0$, $r_0$. A corresponding CP diagram can be constructed as follows (cf.\ Figs.\ \ref{fig:1}-\ref{fig:2}):
\begin{figure}
\center
\epsfig{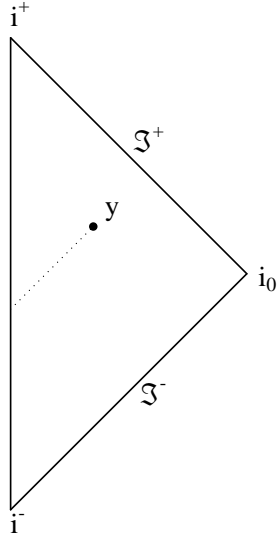}
\caption{CP diagram of a single VBH; the point $y$ corresponds to $u=u_0$, $r=r_0$ in (\ref{ds2})}
\label{fig:1}
\end{figure}
\blist
\item Take Minkowski spacetime (or whatever corresponds to the geometry implied by the chosen boundary conditions on the auxiliary fields $X,X^a$) and draw $N$ different points in its CP diagram; see left diagram of Fig.\ \ref{fig:2}.
\item Draw $N$ copies of this CP diagram and add one light light cut to each (always ending at a different point); remove the other $N-1$ points; the line element is given by (\ref{ds2}); see middle diagram of Fig.\ \ref{fig:2} (note: each of these CP diagrams is equivalent to the one depicted in Fig.\ \ref{fig:1} with varying endpoint $y$).
\item Glue together all CP diagrams at $\scri^\pm$ and $i^0$ (which is a common boundary to all these diagrams); see right diagram of Fig.\ \ref{fig:2}.
\item Take the limit $N\to\infty$.
\elist 
One should not take the effective geometry at face value -- this would be like 
over-interpreting the role of virtual particles in a loop diagram. It is a 
nonlocal entity and one still has to ``sum'' (read: integrate) over all 
possible geometries of this type in order to obtain the nonlocal vertices and 
the scattering amplitude. Nonetheless, the simplicity of this geometry and the 
fact that all possible configurations are summed over are nice features of 
this picture. Because all VBH geometries coincide asymptotically the boundaries of the diagram, $\scri^\pm$ and $i^0$, behave in a classical way\footnote{Clearly the imposed boundary conditions play a crucial role in this context. They produce effectively a fixed background, but only at the boundary.} (thus enabling one to construct an ordinary Fock space like in fixed background QFT), but the more one zooms into the geometry the less classical it becomes.

The coherent sum over all virtual BHs yields a finite S-matrix for ingoing modes with momenta $q,q'$ and outgoing ones $k,k'$,
\begin{equation}
T(q, q'; k, k') = -\frac{i\kappa\delta\left(k+k'-q-q'\right)}{2(4\pi)^4 
|kk'qq'|^{3/2}} E^3 \tilde{T}\,,
\end{equation}
with the total energy $E=q+q'$, $\kappa=8\pi G_N$, 
\begin{eqnarray}
\tilde{T} (q, q'; k, k') := \frac{1}{E^3}{\Bigg [}\Pi \ln{\frac{\Pi^2}{E^6}}
+ \frac{1} {\Pi} \sum_{p \in \left\{k,k',q,q'\right\}}  
\nonumber \\
p^2 \ln{\frac{p^2}{E^2}} {\Bigg (}3 kk'qq'-\frac{1}{2}
\sum_{r\neq p} \sum_{s \neq r,p}\left(r^2s^2\right){\Bigg )} {\Bigg ]}\,,
\end{eqnarray}
and the momentum transfer function $\Pi = (k+k')(k-q)(k'-q)$. The interesting 
part of the scattering amplitude is encoded in the scale independent factor 
$\tilde{T}$. The forward scattering poles occurring for $\Pi=0$ should be noted.

Physically the s-waves of the massless Klein-Gordon field are scattered on their own gravitational self-energy. By rearrangement of the outer legs also a decay of an ingoing s-wave into three outgoing ones is possible.

\section{Deformed SS BH}

Inspired by recent work on deformed Lorentz symmetry\cite{Magueijo:2001cr} (the different original proposal has been presented a decade before\cite{Lukierski:1991pn}) Mignemi constructed deformed dilaton gravity models\cite{Mignemi:2002hd}. 
However, it was shown\cite{Grumiller:2003df} that either the deformation is completely trivial (i.e.\ the metric is unchanged) or the resulting ``metric'' has pathological features of non-invariance. Thus, deformations implemented in this way do not seem to be very useful.

\section{Dynamically deformed SS BH}

\begin{figure}
\centering
\epsfig{file=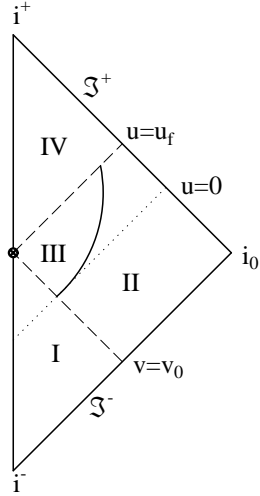,width=0.28\linewidth}
\caption{A sketchy conformal diagram depicting the evaporation of a SS BH}
\label{fig:CP}
\end{figure}

A different implementation of deformations has been proposed recently\cite{Grumiller:2003hq}: exploiting the result on the most general consistent deformation (in the sense of Barnich and Henneaux\cite{Barnich:1993vg}) of 2d BF-theory\cite{Izawa:1999ib} together with assumptions on the asymptotics, the global structure and boundedness of the asymptotic matter flux an evolution equation for the deformation parameter has been established. For the SS BH an attractor solution has been found corresponding to flat spacetime. The evolution could be described in this manner only down to a mass of $M_{\rm Planck}/2$. This remnant mass supposedly is emitted in a final ``thunderbolt''. The CP diagram (cf.\ Fig.\ \ref{fig:CP}) 
has similarities to the one of the Frolov-Vilkovisky model\cite{Frolov:1981mz}: at $v=v_0$ a shock wave creates the BH (thus region I is Minkowski spacetime); at $u=0$ the Hawking process commences (thus region II is Schwarzschild spacetime); at $u=u_f$ the thunderbolt is emitted (in region III dynamical deformation occurs); region IV is the remnant geometry (flat spacetime).

\section*{Acknowledgements}

This work has been supported by projects P-15015, P-14650 of the 
Austrian Science Foundation (FWF) and by CBPF. I am very grateful to 
J.A.~Helayel-Neto and M.~Schweda for crucial help with organizational issues 
and to G.~Amelino-Camelia for the kind invitation. Most of the results 
presented in this talk have been obtained in a fruitful collaboration with 
W.~Kummer and D.V.~Vassilevich and I am deeply indebted to them for numerous 
enlightening discussions.


\end{document}